\documentstyle[prb,aps]{revtex}

\begin{document}

\twocolumn[\hsize\textwidth\columnwidth\hsize\csname
@twocolumnfalse\endcsname
\title{Suppression of matching field effects by splay and pinning energy\\
dispersion in YBa$_2$Cu$_3$O$_7$ with columnar defects}
\author{D. Niebieskikwiat, A. Silhanek, L. Civale, and G. Nieva}
\address{Comisi\'{o}n Nacional de Energ\'{\i}a At\'{o}mica-Centro At\'{o}mico Bariloche\\
and Instituto Balseiro, 8400 Bariloche, Argentina}
\author{P. Levy}
\address{Comisi\'{o}n Nacional de Energ\'{\i}a At\'{o}mica-Departamento de F\'{\i}sica,
Av. Libertador 8250, 1429 Buenos Aires, Argentina}
\author{L. Krusin-Elbaum}
\address{IBM T. J. Watson Research Center, Yorktown Heights, New York 10598}

\maketitle

\begin{abstract}

We report measurements of the irreversible magnetization $M_i$ of
a large number of YBa$_2$Cu$_3$O$_7$ single crystals with columnar
defects (CD). Some of them exhibit a maximum in $M_i$ when the
density of vortices equals the density of tracks, at temperatures
above $40K$. We show that the observation of these {\it matching
field effects} is constrained to those crystals where the
orientational and pinning energy dispersion of the CD system lies
below a certain threshold. The amount of such dispersion is
determined by the mass and energy of the irradiation ions, and by
the crystal thickness. Time relaxation measurements show that the
matching effects are associated with a reduction of the creep
rate, and occur deep into the collective pinning regime.

\end{abstract}


\vskip2pc] \narrowtext

\section{INTRODUCTION}

Vortex dynamics in high temperature superconductors (HTSC) with
columnar defects (CD) depends on a number of variables such as the
density and angular distribution of the CD, the intensity and
orientation of the applied magnetic field ${\bf H}$, and the
temperature $T$. The complexity of these systems results in the
existence of a rich variety of pinning and creep
regimes.\cite{nel-vin,blatter}

The simplest case to model\cite{nel-vin,blatter} is that of
identical and perfectly parallel CD and ${\bf H}$ parallel to
them. At low $T$, and for $H$ much smaller than the {\it matching
field} $B_{\Phi}$ (the dose-equivalent field at which the
densities of vortices and CD are the same), vortex-vortex
interactions can be neglected and each vortex is individually
pinned to an individual track. For $H>B_{\Phi}$ there are more
vortices than defects and pinning becomes collective. As $T$
increases, thermal fluctuations progressively reduce the effective
pinning energy of the CD, thus the vortex-vortex interactions turn
more significant and the boundary separating the individual from
the collective regimes, the so-called {\it accommodation field}
$B_{a}\left( T\right)$, decreases.\cite{accommod}

An interesting situation occurs for $H \sim B_{\Phi}$. At low $T$
a Mott insulator phase is predicted.\cite{nel-vin,blatter} The
fingerprint of this phase is the infinite elastic compression
modulus $C_{11}$, that results in a constant induction field $B$
(fixed density of vortices) over a finite range of $H$. The
dynamics is also influenced by the matching condition. As all the
CD are occupied, a pinned vortex has no energetically convenient
places to jump into, with the consequent reduction of the creep
rate. The Mott phase has indeed been observed in time relaxation
experiments at very low $T$, by Beauchamp {\it et
al.}\cite{beauchamp} in YBa$_2$Cu$_3$O$_7$ (YBCO) crystals and by
Nowak {\it et al.}\cite{nowak} in Tl:2201 crystals.

At high $T$ the wandering of the vortex lines precludes their
localization into individual CD. Although this effect should
inhibit the appearance of the Mott insulator
phase,\cite{nel-vin,blatter} some reduction in the vortex mobility
is still expected due to the absence of empty tracks. However,
many studies of vortex pinning by CD in HTSC have failed to show
any evidence of {\it matching effects} at high temperatures. This
status was modified by a recent study performed by Mazilu {\it et
al.}\cite{mazilu} on YBCO thick films (thickness $\delta \sim 1\mu
m$) with CD $\parallel c$-axis. For ${\bf H} \parallel$ CD, they
observed that the transport critical current had a broad peak at
$H \sim B_{\Phi}$, at temperatures as high as 75K.

In this work we report the observation of matching effects due to
CD introduced by heavy ion irradiation in YBCO single crystals. We
observe such effects at high temperatures, deep into the
collective pinning regime. For crystals with CD in different
crystallographic orientations, we find that for ${\bf H}\parallel$
tracks, the irreversible magnetization $M_i(H)$ exhibits a local
maximum at $H \sim B_{\Phi}$. That maximum is associated with a
local minimum in the normalized time relaxation rate $S=-d(\ln
M_{i})/d(\ln t)$. We show that the appearance of these matching
effects requires a narrow angular distribution (small splay) and a
small pinning energy dispersion of the CD. These conditions impose
a maximum track length (given by the sample thickness and the
irradiation angle) that depends on the mass and energy of the
irradiation ions.

\section{EXPERIMENTAL}

We have observed matching effects in four YBCO single crystals.
For comparison, we show analogous measurements in several other
YBCO crystals that do not exhibit such effects. A group of
crystals was grown at the Centro At\'{o}mico Bariloche\cite{paco} and
was irradiated with $300MeV$ $Au^{24+}$ ions at the Tandar
facility in Buenos Aires, Argentina.\cite{physicac} Another group
was grown at the T.J. Watson Research Center of
IBM.\cite{holtzberg} Some of them were irradiated\cite{arresting}
with $1.08GeV$ $Au^{23+}$ at the TASCC facility in Chalk River
Laboratories, Canada, and the rest\cite{leo-CD} with $580MeV$
$Sn^{30+}$ at the Holifield accelerator, Oak Ridge, USA. The
crystal thicknesses $\delta$, the matching fields $B_{\Phi}$ and
the angle $\Theta_D$ between the incident beam and the $c$-axis
are summarized in table I.

Measurements of dc magnetization were made in two commercial
superconducting quantum interference device (SQUID) magnetometers,
both equipped with $50kOe$ magnets. The irreversible magnetization
$M_i$ (proportional to the persistent current density $J$ via the
critical state model) was determined from $M(H)$ loops. The
magnetic field was always applied parallel to the irradiation
direction. In crystals where $\Theta_D \neq 0$, both components of
${\bf M}_i(H)$ were recorded using the two sets of pick up coils
(longitudinal and transverse), and the data were processed in the
way previously described.\cite{silhanek} Relaxation measurements
of $M_i$ were performed on the field decreasing branch of the
hysteresis loop over periods of 2 hours.

\section{RESULTS}

\subsection{Sample thickness influence on matching effects}

In Fig. 1(a) we show $M_{i}$ as a function of $H$ for crystal A1
at several temperatures between $40K$ and $75K$, for ${\bf
H}\parallel$ tracks ($\Theta_D \approx 57^{\circ}$). These curves
show a clear maximum at fields $H_m(T)$ close to $B_{\Phi}$. This
{\it matching effect} is very similar to that found by Mazilu {\it
et al.}\cite{mazilu} in transport measurements of critical current
in YBCO thick films with ${\bf H}\parallel$ CD$\parallel c$-axis.
The temperature dependence of $H_m/B_{\Phi}$ for both our crystal
and a thick film with $B_{\Phi} \sim 3.9T$, taken from Ref. 6, is
shown in Fig. 2. Several features reinforce the similarity between
the results of that study and our data. First, at the lowest
temperatures the maximum occurs slightly above the matching field
$B_{\Phi}$. Second, the field $H_m$ slowly decreases with
increasing $T$. Third, the maximum disappears at low temperatures,
below $40K$ in our case.

Many measurements of $M_i$ in YBCO crystals with CD, in the same
field and temperature ranges of Fig. 1, can be found in the
literature. Usually, the maximum shown in Fig. 1(a) is not
observed. The question is why these matching effects are visible
in some cases and not in others. Clearly, it is not due to a
structural characteristic of the thick films, as they are also
visible in a crystal. The orientation of the CD is not relevant
either. A distinctive characteristic of crystal A1 is that it is
unusually thin, $\delta \approx 4.1\mu m$. The thick films
measured by Mazilu {\it et al.}\cite{mazilu} are of course even
thinner, $\delta \sim 1\mu m$. In contrast, most YBCO crystals
with CD reported in previous studies have typical thicknesses
between $10\mu m$ and $30\mu m$. This observation suggests that
the appearance of matching effects may be restricted to thin
samples.

To test this hypothesis, we have collected $M_i(H)$ data for a
representative group of 13 additional crystals. We have included
crystals irradiated with ions of different mass and energy, at a
variety of $B_{\Phi}$ and $\Theta_D$, as indicated in table I. In
all cases ${\bf H} \parallel$ CD. Figure 3 shows $M_i(H)$ curves
for 10 crystals. None of them show any evidence of matching
effects. For clarity only data at $T=60K$ are shown, but in all
cases the maximum is also absent at other temperatures within the
range $40K \leq T \leq 80K$. In contrast, the $M_i(H)$ curves of
crystals B1 and B2, shown in Figs. 4(a) and (b) respectively, do
exhibit a clear maximum in the vicinity of $B_{\Phi}$. Again in
these two crystals $H_m$ decreases slowly with temperature.
Finally, in crystal B3, shown in Fig. 4(c), a small structure in
$M_i(H)$ near $B_{\Phi}$ just suggests the existence of minor
matching effects. In the three crystals (B1, B2 and B3) the
matching effects disappear below $40K$.

Inspection of Figs. 1 and 4 show that, at $T \sim 40K$, the ratio
$H_m/B_{\Phi}$ is $\sim 1.1$ for crystal A1, while it is only
$\sim 0.75$ for B1 and $\sim 0.85$ for B2. This difference may be
due to clustering of the tracks.\cite{accommod} As $B_{\Phi}$
increases, also does the probability that two or more CD are so
close together that they act as a single one. The result is an
``effective" matching field lower than the nominal $B_{\Phi}$. For
crystal B1, the effective tracks' density was found\cite{accommod}
to be $\sim 0.7 B_{\Phi}$, while for a crystal with a dose similar
to A1 the result was $\sim 0.9 B_{\Phi}$. Thus, there is a
reasonable agreement between both results, indicating that $H_m$
is in all cases slightly higher than the {\it effective} matching
field.

To some extent data in Figs. 3 and 4 reinforce the idea that
matching effects are associated with thin samples, as crystals B1
and B2 are among the thinnest in the group. However, it is clear
that the correlation is far from perfect; crystals A2, A3 and B5,
which are as thin as B1 and B2 or even thinner, show no maximum at
$B_{\Phi}$. Another exception is crystal B3, which is rather thick
and shows at least a hint of the effect.

The simplest connection between sample thickness and matching
effects could be related to the inhomogeneity of the internal
field $B$. It is natural to expect that matching effects should be
visible over a finite field range $\Delta B_{me}$ around
$B_{\Phi}$. On the other hand, in the critical state of a thin
superconducting slab in a transverse field configuration, $B$ has
a total variation\cite{clem-brandt} $\Delta B\sim
\left(4\pi/c\right) J \delta/2$ between the center and the border
of the sample. Therefore, the required conditions for observation
of matching effects can be satisfied over the whole sample only if
$\Delta B < \Delta B_{me}$. Otherwise, the maximum in $M_i (H)$
will be washed away.

We can estimate $\Delta B$ at $T=60K$ and in the proximity of
$B_{\Phi}$ for all the crystals. It ranges from $\Delta B \sim
0.12 kG$ for crystal A1 to $\Delta B \sim 0.6 kG$ for crystals C4
and C5. An estimate of $\Delta B_{me}$ is given by the width of
the maximum in $M_i(H)$. From Figs. 1 and 4 we see that $\Delta
B_{me}$ is similar in crystals A1, B1 and B2, and in all cases it
is well above $10 kG$. Thus, we conclude that even in the thickest
crystals $\Delta B \ll \Delta B_{me}$, so the inhomogeneity in the
internal field cannot be the reason for the absence of matching
effects.

\subsection{Energy and angular dispersion of the CD}

Sample thickness influences pinning by CD in a more indirect way,
as it affects the morphology and the disorder of the tracks. When
a sample is irradiated, the heavy ions arrive to the surface with
an extremely narrow distribution of energy and orientations. Thus,
all the CD are initially identical and exactly parallel. But as
they penetrate deeper into the material, their energy decreases
rapidly due to the very large electronic stopping power. As a
result, the diameter of the tracks initially decreases as a
function of depth, then becomes oscillatory and eventually the
tracks turn discontinuous as the ions approach their penetration
range.\cite{studer} Consequently, there is a certain dispersion in
the pinning energy of the CD. In addition, the scattering with the
atomic cores of the material (associated with the small but
nonzero nuclear stopping power) deviates the incident ions from
the original direction. The cumulative effect of these nuclear
interactions produces\cite{arresting,wheeler} an angular
dispersion (splay) of the tracks that grows with depth, first
slowly and then dramatically near the ion penetration range. These
effects have been clearly demonstrated in several transmission
electron microscopy (TEM) studies.\cite{arresting,studer,wheeler}

The above considerations indicate that, in all cases, real CD
introduced by swift heavy-ion irradiation contain some amount of
splay and pinning energy dispersion. For given irradiation
conditions, the distributions of pinning energies and orientations
are wider for thicker samples. We will now argue that this larger
dispersion in the tracks precludes the observation of matching
effects in thick crystals.

At this point we need a more quantitative measure of the amount of
disorder in the system of CD. We first note that the relevant
parameter is not the thickness of the crystal, but rather the
length of the tracks, $l_D= \delta / \cos \Theta_D$, which are
also listed in table I. We see that the shorter $l_D \sim 7.5 \mu
m$ still corresponds to crystal A1. The $l_D$ of crystals B1 and
B2 still rank among the shortest, and the discrepancy of B5 is
solved, as due to the large $\Theta_D$ we have a large $l_D \sim
27 \mu m$ for this case.

Splay and energy dispersion as a function of depth depend in a
complex way on the ion mass and initial
energy.\cite{studer,wheeler} We look for a simpler description
where disorder is characterized by a single parameter. In Ref. 10
the median radial angle $\alpha_{SP}$ of the angular distribution
of CD was determined as a function of depth in YBCO crystals, for
CD produced by $1.08GeV$ $Au^{23+}$ and $580MeV$ $Sn^{30+}$. The
determination was based on TRIM numerical calculations that
coincided very well with direct measures of $\alpha_{SP}$ at
selected depths from TEM images. We have repeated those TRIM Monte
Carlo calculations and extended them to the $300MeV$ $Au^{24+}$
case. If our assumption is correct, matching effects will be
erased if $\alpha_{SP}$ is large enough. Thus, to quantify the
splay of the CD we have chosen the largest $\alpha_{SP}$ in each
crystal, which occurs at the back end of the tracks.

In Fig. 5 we plotted $\alpha_{SP}$ at $l_D$ for all our crystals.
Note that the data points are separated in 3 groups corresponding
to each type of irradiation. An additional data point
corresponding to the thick films of Ref. 6 is also included. It is
unmistakably clear that there is a threshold value of $\alpha_{SP}
(l_D) \approx 3.4^{\circ}$ above which the matching effects
disappear. All samples with $\alpha_{SP} (l_D)$ well below the
threshold (B1, B2 and the thick films) exhibit clear matching
effects. None of the nine crystals with $\alpha_{SP} (l_D)$ well
above the threshold show any hint of it. Finally we have three
crystals, each one of them irradiated in different conditions,
with almost exactly the same $\alpha_{SP} (l_D) \approx
3.4^{\circ}$. One of them shows a clear matching effect (A1, see
Fig. 1); another one shows just a minor hint [B3, Fig. 4(c)]; and
the third one shows no effect (C1, Fig. 3).

Figure 5 shows that, using a single parameter, we have been able
to ascertain under what conditions matching effects are observable
in YBCO with CD. We do not claim that $\alpha_{SP} (l_D)$ is the
only or even the best quantifier of the track´s disorder; it is
just a simple,  reasonable and convenient one. An implicit
assumption, for instance, is that splay and pinning energy
dispersion are strongly correlated and thus can be characterized
by a single number. A more elaborated analysis could produce a
better quantifier, but we should emphasize that our description
successfully describes the behavior of our 14 crystals and all the
films of Ref. 6, with no exceptions.

\subsection{Relaxation studies}

We have observed the maximum at $H_m$ in temperatures ranging from
$40K$ up to as high as  $80K$, very close to the irreversibility
line.\cite{leo-irr} This result is somewhat surprising, since the
Mott phase is only expected at low $T$.\cite{nel-vin,blatter} It
is true that the broad peak in $M_i(H)$ seen in our crystals and
in the films of Ref. 6 (where it was described as a {\it vestige}
of the Mott phase) is a feature far less dramatic than the
Meissner-like response of the Mott insulator. But on the other
hand, the physical origin is clearly the same: The maximum in the
pinning efficiency of the CD at $B \sim B_{\Phi }$ occurs because,
being all the vortices pinned and all the tracks occupied, there
are no energetically convenient places for a vortex to move on
from its initial position.

If the above picture is correct, the maximum in $M_i(H)$ should be
accompanied by a decrease of the creep rate. This is precisely the
feature used by Beauchamp {\it et al.}\cite{beauchamp} and by
Nowak {\it et al.}\cite{nowak} to identify the Mott insulator
phase at very low $T$. Fig. 1(b) shows the normalized relaxation
rate $S$ as a function of ${\bf H} \parallel$ CD at $T=60K$ for
crystal A1. For $H\sim B_{\Phi }$ a local minimum appears, thus
confirming that matching fields effects at these high temperatures
are due to a reduction of the creep processes.

It is important to realize that in the temperature range of our
observations the vortex system is in the collective pinning
regime.\cite{accommod} Crystal B1 is the same one investigated in
Ref. 3. It was shown there that the {\it accommodation field}
$B_{a}\left( T\right) $ (the boundary between individual and
collective pinning) goes to zero at the {\it depinning
temperature} $T_{dp}\approx 40K$, thus for $T>T_{dp}$ the pinning
is collective for all values of $H$. The abrupt collapse of
$B_{a}\left( T\right) $ at $T_{dp}\approx 40K$ was observed not
only in B1 but also in all the other crystals reported there
(irradiated in all cases with $1.08GeV$ $Au^{23+}$), regardless of
$B_{\Phi }$. Recent results\cite{dario} indicate that $T_{dp}$ for
crystals irradiated with $580MeV$ $Sn^{30+}$ and $300MeV$
$Au^{24+}$ is also very similar.

To our knowledge, a determination of $B_{a}(T)$ for tracks'
orientations other than the c-axis was not available. Thus, to
check whether the large $\Theta_D$ makes any difference in this
respect, we determined $B_{a}(T)$ for crystal A1. To that end we
measured the normalized relaxation rate $S$ as a function of $T$
at several ${\bf H} \parallel$ CD, as shown in the inset of Fig.
2. The maxima in these curves, which indicate the onset of the
collective pinning regime, have been used\cite{accommod} as a
signature of $B_{a}(T)$. The collective regime becomes completely
developed when $S(T)$ reaches the {\it plateau} at higher
temperatures.\cite{physicac,dario} The temperature range in
between the maxima and the {\it plateau} (i.e., the region where
$S$ decreases with $T$) is a transition zone where individual and
collective excitations coexist.\cite{dario} The {\it accommodation
field} and the transition zone are included in the complete
dynamic phase diagram shown in the main frame of Fig. 2. This
diagram again shows that $T_{dp}\approx 35K$, and it clearly
demonstrates that also in crystal A1 the vestiges of the Mott
insulator phase (the $H_m(T)$ line) lie well inside the collective
pinning regime.

Within this regime and at fixed $T$, the critical current should
have a $1/H$ dependence.\cite{nel-vin,blatter} Therefore, a
monotonically increasing $S(H)$ is also expected. This is indeed
the observed behavior in Fig. 1(b), with the local minimum at
$B_{\Phi}$ mounted on the increasing curve. This feature once
again shows the presence of the matching effects in the collective
pinning regime.

\section{DISCUSSION}

\subsection{Dispersion-induced slow-down of creep}

The two main results of our study are that matching effects occur
deep into the collective pinning regime, and that these effects
are destroyed by sufficiently large splay and dispersion of
pinning energy in the system of CD. We will now discuss the second
result, and in the next subsection we will address the first one.

According to the theoretical models,\cite{nel-vin,blatter} the
main features of the pinning diagram are expected to be robust
with respect to the energy dispersion and splay: Pinning by CD in
HTSC should still be individual and strong at low T and H, and
should become collective and weaker above $B_a(T)$, as either T or
H increase. Experimental results confirm that
expectation,\cite{accommod,physicac,leo-CD,dario} as the basic
pinning behavior is similar in all samples, in spite of the fact
that they contain different amounts of dispersion in the CD
system. In contrast, the dynamic at current densities well below
$J_c$ is strongly influenced by the dispersion in the
CD.\cite{arresting,dario} This indicates that the link between the
amount of dispersion in the CD and the observation of matching
effects must be related to the influence of the dispersion on the
creep processes.

In the single vortex pinning regime, time relaxation at the early
stages takes place\cite{nel-vin,blatter} via nucleation and
expansion of {\it half loops}. As $J$ decreases the size of the
critical nucleus grows and eventually reaches the nearest CD.
Further relaxation proceeds by spreading of the resulting {\it
double kink} vortex excitations. Ideally, if energy dispersion and
splay are not taken into account, there is no barrier for the
expansion of a double kink critical nucleus, and $J$ should
decrease very rapidly.\cite{nel-vin,blatter} In the collective
pinning regime the creep mechanisms are somewhat different and
less explored theoretically. However, vortex bundles are also
expected to relax via collective double kinks,\cite{blatter} whose
expansion is again unimpeded in the absence of splay and energy
dispersion.

Both splay and energy dispersion of the CD will arrest the
expansion of double kinks, since both reduce the number of sites
with equivalent energy available for the hopping and spreading
processes. As a result, the $J$ determined from magnetization
measurements is larger than it should be in the absence of
dispersion. The idea of topological constrains on vortex hopping
was first discussed by Hwa {\it et al.}\cite{hwa} It is indeed
established experimentally \cite{splay} that a certain amount of
splay enhances $M_i$ in YBCO. In particular, by comparing YBCO
crystals of different thicknesses, irradiated with different ions
and energies, it has been shown that the splay reduces the creep
rate.\cite{arresting}

Energy dispersion makes the expansion of double kinks
energetically unfavorable in the limit $J \rightarrow
0$.\cite{nel-vin,blatter} Double kink excitations are then
substituted by {\it superkinks}, whose associated time relaxation
(the so called {\it variable range hopping} regime) is much
slower. We have recently demonstrated\cite{dario} that fast
relaxation by double kinks does occur in YBCO crystals, and that
the crossover to the superkinks regime and the associated
slow-down of the creep takes place at a current density $J_{VRH}$
which is proportional to the energy dispersion.

The above discussion leads us to propose the following scenario.
In those samples where splay and energy dispersion are small, time
relaxation is fast and the overall measured $M_i$ is low. However,
near the matching condition creep slows down due to the absence of
available sites. As a result, the $M_i$ measured at a given time
is higher around $B_{\Phi}$ than in the rest of the field range,
thus producing the observed maximum. In contrast, the large amount
of splay and energy dispersion in thick samples limits the
expansion of double kinks at all fields. As a consequence, the
creep rate becomes lower in the whole field range, and the creep
reduction near $B_{\Phi}$ becomes negligible or entirely absent.
In these conditions, the overall $M_i$ is high and the distinct
maximum at the matching condition disappears.

Finally, the weak decrease of $H_m$ with temperature could be
related to the small but nonzero dispersion of pinning energies.
The effectiveness of the CD is strongly reduced with increasing
$T$ due to the entropic smearing.\cite{accommod} Therefore, some
weak CD can pin vortices at low temperatures but cannot hold them
pinned at high $T$, thus slightly reducing the effective matching
field.

\subsection{Matching effects in the collective pinning regime}

Previous numerical simulations\cite{wengel} of matching effects
within the collective pining regime have been able to predict some
effects of the Mott phase at high temperatures, but only in the
case $\lambda /d\leq 1$, where $\lambda $ is the penetration
length and $d=\sqrt{\Phi _{0}/B_{\Phi }}$ the average distance
between CD. This inequality results from the condition that the
vortex-vortex interactions have to be of short range as compared
with the distance between tracks (that equals the distance between
flux lines at $B=B_{\Phi }$). However, in our case $\lambda
\approx 1400\AA $ and $d$ ranges from $190\AA$ for crystal B2 to
$300\AA$ for crystal A1, which gives $\lambda/d \geq 4.7$ in all
cases.

Krauth {\it et al.}\cite{krauth} have studied the problem of 2D
bosons in a disordered environment, which is analogous to the
problem of flux lines in the presence of CD.\cite{nel-vin} Through
Monte Carlo numerical simulations they have found that the Mott
insulator phase could be present up to the transition to the Bose
glass phase. But again, they have used only on-site repulsion,
which is equivalent to the condition $\lambda /d<1$, i.e.,
negligible vortex-vortex interactions.

More recently, Sugano {\it et al.}\cite{sugano} have performed
Monte Carlo numerical simulations of pancake vortices in the much
more anisotropic Bi$_2$Sr$_2$CaCu$_2$O$_8$ system, with CD
parallel to the c-axis. Neither splay nor pinning energy
dispersion were considered. The main result of that study is the
observation of a field-driven discontinuous transition in the
trapping rate of the pancakes to the CD, at a field $\sim
B_{\Phi}/3$, that is accompanied by a large jump in the interlayer
coherence. This result is in agreement with recent Josephson
Plasma resonance experiments\cite{sato} and c-axis resistivity
measurements\cite{morozov} in the liquid phase, and is also
consistent with several experimental observations of anomalies in
the solid phase.\cite{vanderbeek} Similarly, anomalies in
$H_{irr}$ and maxima in $M_i(H)$ have been
observed\cite{leo-CD,leo-irr} in the YBCO system, in the field
range $\sim B_{\Phi}/3$ to $\sim B_{\Phi}/2$. An additional result
of these simulations, particularly relevant to our present work,
is the observation\cite{sugano} of a high temperature "subanomaly"
at the matching field $\sim B_\Phi$, thought to be a {\it remnant}
of the low-temperature Mott insulator phase. This subanomaly is
accompanied by a sudden increase in the vortex trapping rate,
i.e., a slow down  of creep, which according to those simulations
is dominated by expansion of {\it double kinks}. All these results
are consistent with our scenario, according to which the maximum
in $M_i(H)$ at $H \sim B_\Phi$ is due to the reduction of the rate
of creep by double kink excitations. Topological vortex
entanglement, forced by splay present in the thicker crystals, as
well as large pinning energy dispersion, erase this matching
subanomaly by constraining the expansion of double kinks at all
fields.

\section{Conclusions}

We have observed matching field effects in the irreversible
magnetization and its time relaxation in YBa$_2$Cu$_3$O$_7$ single
crystals with columnar defects. We have demonstrated that a
necessary condition for the appearance of these effects is a low
level of angular and energy dispersion in the CD system. To
achieve this situation, an adequate combination of thin samples
and high irradiation energy is required. Large dispersion
precludes the appearance of matching effects by slowing down the
creep processes over the whole field range. We propose the value
of the splay at the back face of the sample, $\alpha_{SP} (l_D)$,
as a convenient parameter to quantify the dispersion.
Surprisingly, these matching effects are observed at high
temperatures, deep into the collective pinning regime, where the
Mott phase is not expected.

\begin{center}
{\small {\bf ACKNOWLEDGMENTS}}
\end{center}

This work was partially supported by ANPCyT, Argentina, PICT 97
No. 01120. A.S. and G.N. would like to thank the CONICET for
financial support.

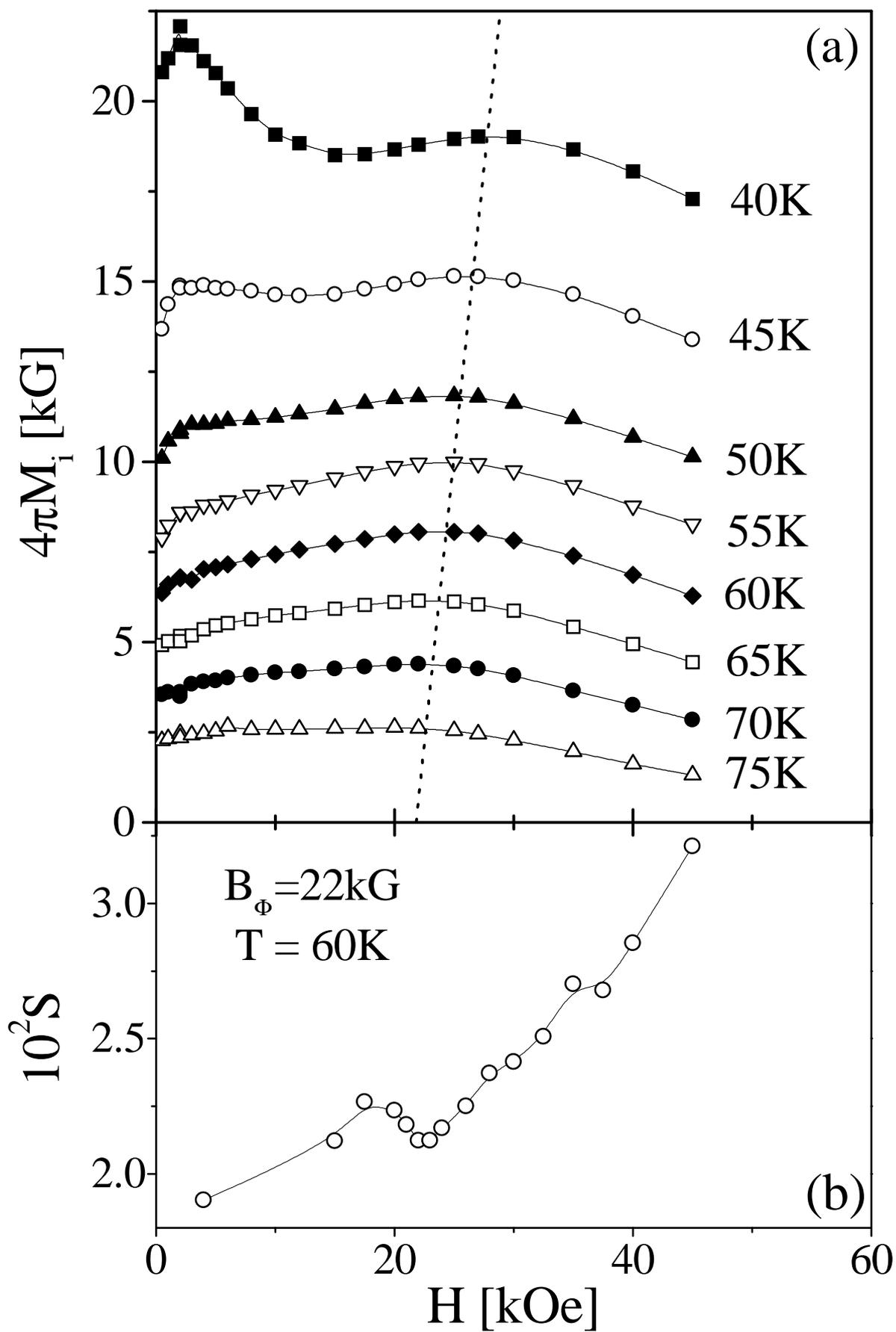
FIG. 1. (a) Curves of irreversible magnetization $M_{i}$ vs
applied field $H$ for crystal A1 at several temperatures for ${\bf
H}\parallel$ CD. The dotted line is a guide to the eye indicating
the position of the maximum $H_m(T)$. (b) Normalized relaxation
rate as a function of $H$ at $T=60K$. The local minimum is
generated by the reduction of the mobility of the flux lines when
$H \sim B_{\Phi}$.

\vspace{0.5cm}

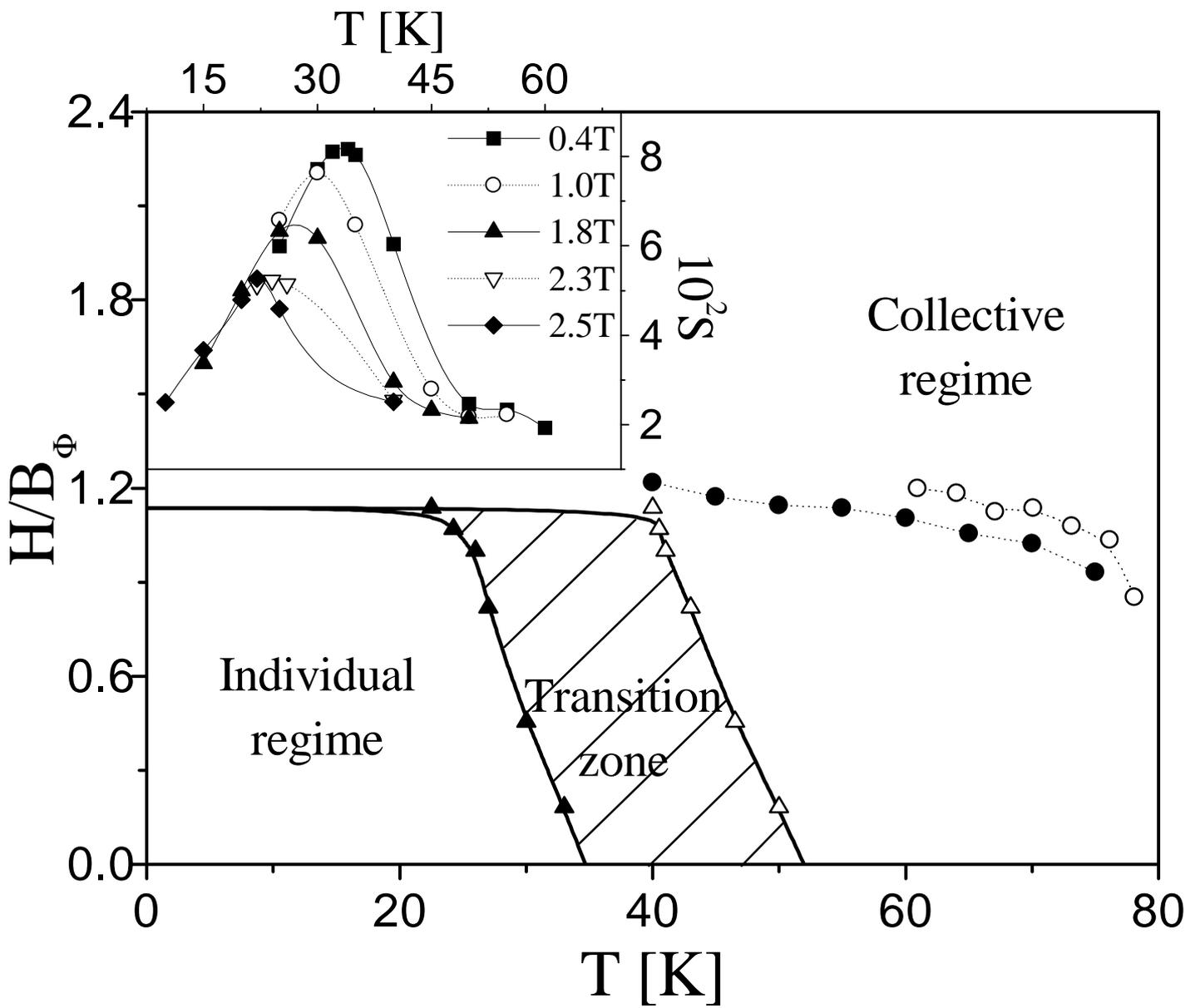
FIG. 2. Dynamic $H$-$T$ phase diagram for crystal A1. The solid
triangles represent the {\it accommodation field} $B_{a}(T)$. The
solid circles indicate the line $H_m(T)/B_{\Phi}$. For comparison,
the $H_m(T)/B_{\Phi}$ data for a thick film of Ref. 6 is also
shown (open circles). Inset: Normalized relaxation rate $S(T)$
curves at several fields. The maximum indicates the onset of the
collective excitations at high $T$.

\vspace{0.5cm}

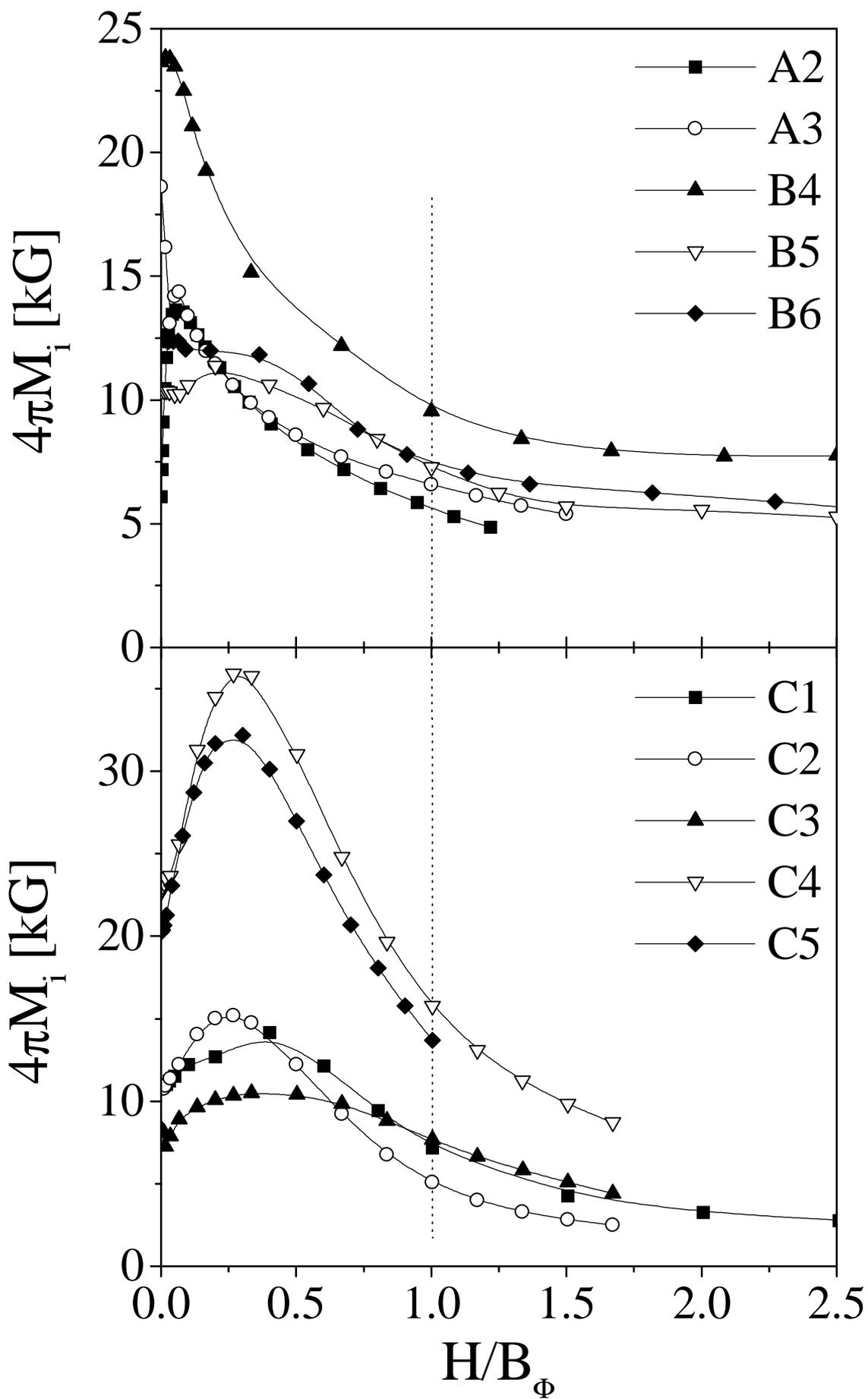
FIG. 3. Irreversible magnetization $M_i$ as a function of
$H/B_{\Phi}$ for several crystals at $T=60K$ and ${\bf
H}\parallel$ CD. None of these samples show any hint of matching
effects at $H \sim B_{\Phi}$.

\vspace{0.5cm}

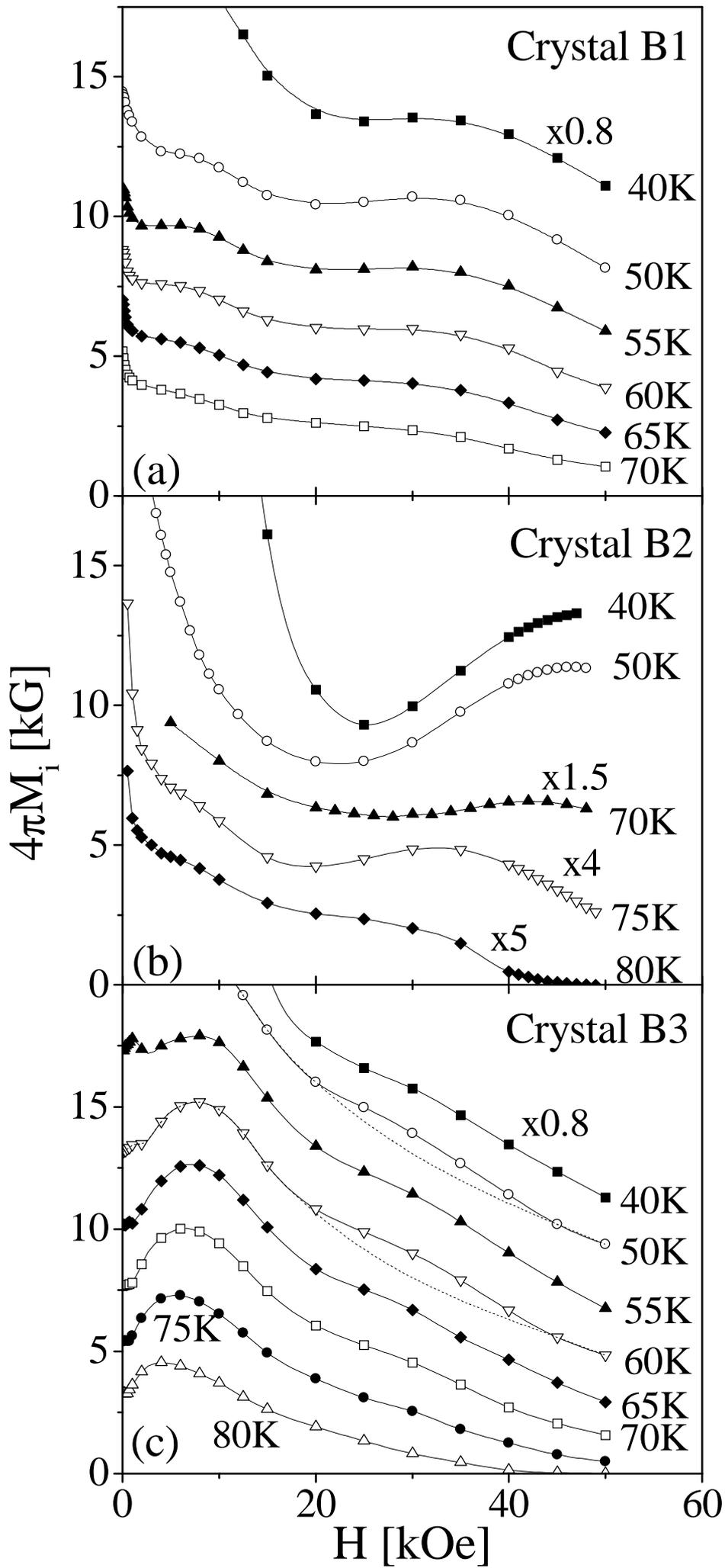
FIG. 4. Irreversible magnetization as a function of $H$ at several
$T$ for three crystals that exhibit matching effects. For clarity,
some curves are multiplied by a numerical factor, as indicated.

\vspace{0.5cm}

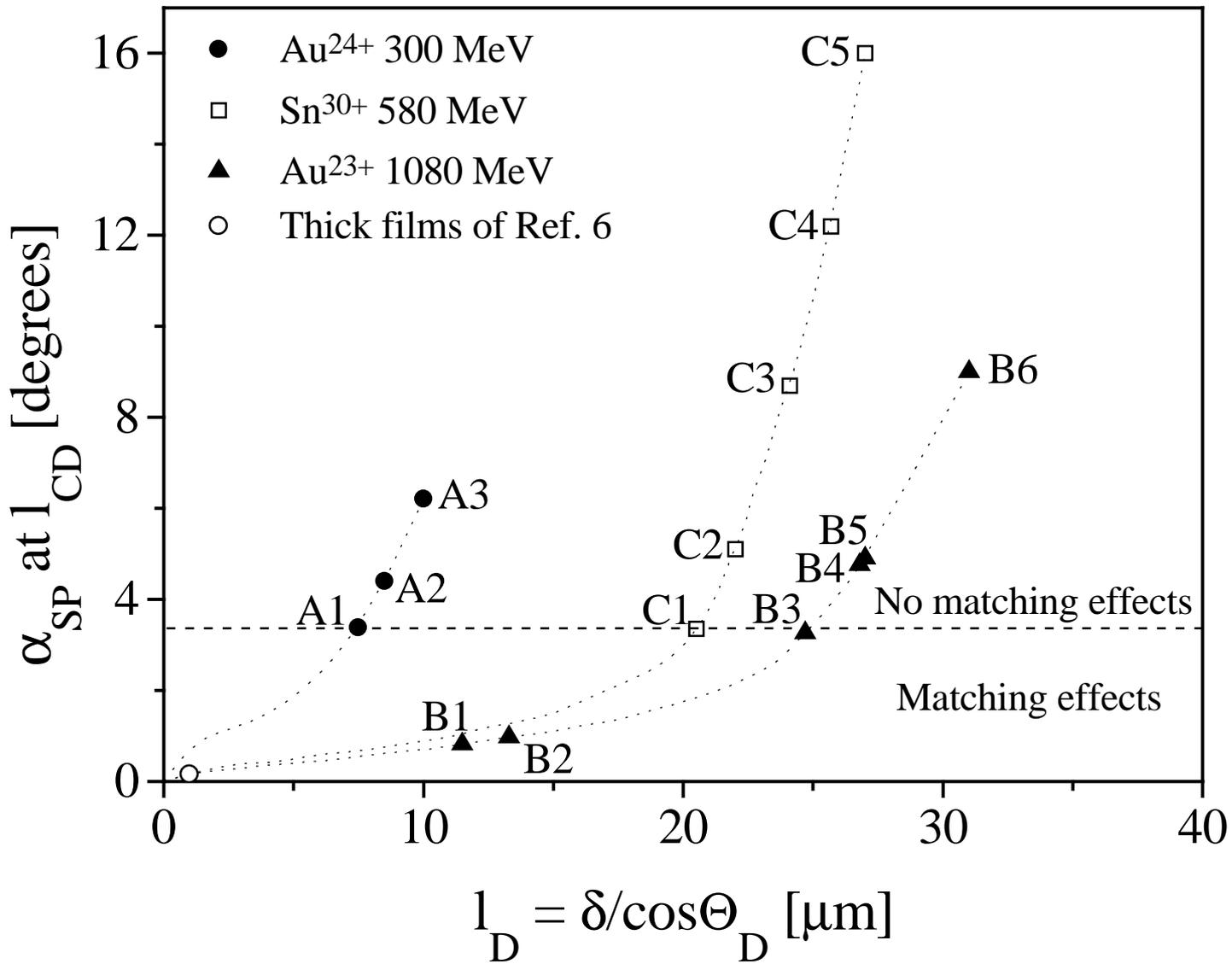
FIG. 5. Median radial angle $\alpha_{SP}$ of the CD at the back
face of the sample as a function of the total tracks' length
$l_D$. Samples with $\alpha_{SP}(l_D)<3.4^{\circ}$ exhibit
matching effects, as opposed to the crystals with
$\alpha_{SP}(l_D)>3.4^{\circ}$.

\newpage

\begin{table}[ht]
\centering \caption{Irradiation and thickness specifications for all the crystals studied. The crystals labeled with an asterisk present matching effects.}
\begin{tabular}[b]{l c c c c c}
Crystal & ion & $B_{\Phi}(T)$ & $\Theta_D$ & $\delta(\mu m)$ &
$\delta/ \cos \Theta_D(\mu m)$ \\
     \hline
A1$^*$ & 300MeV Au$^{24+}$ & 2.2 & $57^{\circ}$ & 4.1 & 7.5 \\

A2 & 300MeV Au$^{24+}$ & 3.7 & $15^{\circ}$ & 8.2 & 8.5 \\

A3 & 300MeV Au$^{24+}$ & 3.0 & $32^{\circ}$ & 8.5 & 10.0 \\

B1$^*$ & 1080MeV Au$^{23+}$ & 4.7 & $0^{\circ}$ & 11.5 & 11.5 \\

B2$^*$ & 1080MeV Au$^{23+}$ & 5.7 & $30^{\circ}$ & 11.5 & 13.3 \\

B3$^*$ & 1080MeV Au$^{23+}$ & 2.4 & $0^{\circ}$ & 24.7 & 24.7 \\

B4 & 1080MeV Au$^{23+}$ & 0.6 & $0^{\circ}$ & 26.8 & 26.8 \\

B5 & 1080MeV Au$^{23+}$ & 1.0 & $65^{\circ}$ & 11.4 & 27.0 \\

B6 & 1080MeV Au$^{23+}$ & 1.1 & $2^{\circ}$ & 31.0 & 31.0 \\

C1 & 580MeV Sn$^{30+}$ & 1.0 & $2^{\circ}$ & 20.5 & 20.5 \\

C2 & 580MeV Sn$^{30+}$ & 3.0 & $2^{\circ}$ & 22.0 & 22.0 \\

C3 & 580MeV Sn$^{30+}$ & 3.0 & $30^{\circ}$ & 20.9 & 24.1 \\

C4 & 580MeV Sn$^{30+}$ & 3.0 & $2^{\circ}$ & 25.7 & 25.7 \\

C5 & 580MeV Sn$^{30+}$ & 5.0 & $2^{\circ}$ & 27.0 & 27.0 \\

\end{tabular}
\end{table}


\begin{references}

\bibitem{nel-vin}  D. R. Nelson and V. M. Vinokur, Phys. Rev. B {\bf 48},
13060 (1993).

\bibitem{blatter}  G. Blatter, M. V. Feigel'man, V. B. Geshkenbein, A.I.
Larkin, and V. M. Vinokur, Rev. Mod. Phys. {\bf 66}, 1125 (1994).

\bibitem{accommod}  L. Krusin-Elbaum, L. Civale, J. R. Thompson, and C.
Feild, Phys. Rev. B {\bf 53}, 11744 (1996).

\bibitem{beauchamp}  K. M. Beauchamp, T. F. Rosenbaum, U. Welp, G. W.
Crabtree, and V. M. Vinokur, Phys. Rev. Lett. {\bf 75}, 3942 (1995).

\bibitem{nowak}  E. R. Nowak, S. Anders, H. M. Jaeger, J. A. Fendrich, W. K.
Kwok, R. Mogilevsky, and D. G. Hinks, Phys. Rev. B {\bf 54}, 12725 (1996).

\bibitem{mazilu}  A. Mazilu, H. Safar, M. P. Maley, J. Y. Coulter, L. N.
Bulaevskii, and S. Foltyn, Phys. Rev. B {\bf 58}, 8909 (1998).

\bibitem{paco}  F. de la Cruz, D. L\'{o}pez, and G. Nieva, Philos.
Mag. B {\bf 70}, 773 (1994).

\bibitem{physicac}  L. Civale, G. Pasquini, P. Levy, G. Nieva, D. Casa, and
H. Lanza, Physica C {\bf 263}, 389 (1996).

\bibitem{holtzberg} F. Holtzberg and C. Feild, Eur. J. Solid State
Inorg. Chem. {\bf 27}, 107 (1990).

\bibitem{arresting}  L. Civale, L. Krusin-Elbaum, J. R. Thompson, R. Wheeler,
A. D. Marwick, M. A. Kirk, Y. R. Sun, F. Holtzberg, and C. Feild,
Phys. Rev. B {\bf 50}, 4102 (1994).

\bibitem{leo-CD} L. Civale, A. D. Marwick, T. K. Worthington, M. A. Kirk,
J. R. Thompson, L. Krusin-Elbaum, Y. Sun, J. R. Clem, and F.
Holtzberg, Phys. Rev. Lett. {\bf 67}, 648 (1991).

\bibitem{silhanek}  A. Silhanek, L. Civale, S. Candia, G. Nieva, G.
Pasquini, and H. Lanza, Phys. Rev. B {\bf 59}, 13620 (1999).

\bibitem{clem-brandt}  E. H. Brandt, Phys. Rev. B {\bf 49}, 9024 (1994);
J. R. Clem and A. Sanchez, {\it ibid} {\bf 50}, 9355 (1994).

\bibitem{studer} See for example F. Studer and M. Toulemonde, Nucl. Instr.
Meth. B {\bf 65}, 560 (1992); V. Hardy {\it et al.}, {\it ibid}
{\bf 54}, 472 (1991); A. D. Marwick {\it et al.}, Proceedings of
the Eighth International Conferenceon Ion Beam Modification of
Materials (1992); D. X. Huang {\it et al.}, Phys. Rev. B {\bf 57},
13907 (1998).

\bibitem{wheeler} R. Wheeler, M. A. Kirk, R. Brown, A. D. Marwick, L. Civale,
and F. Holtzberg, in Phase Formation and Modification by Beam-Solid
Interactions, edited by G. S. Was, L. E. Rehn, and D. M. Follstaedt
(MRS Symposium Proceedings - Vol. 235, Pittsburgh, 1992), p. 683.

\bibitem{leo-irr} L. Krusin-Elbaum, L. Civale, G. Blatter, A. D. Marwick,
F. Holtzberg, and C. Feild, Phys. Rev. Lett. {\bf 72}, 1914
(1994).

\bibitem{dario}  D. Niebieskikwiat, L. Civale, C. A. Balseiro, and G. Nieva,
Phys. Rev. B {\bf 61}, 7135 (2000).

\bibitem{hwa}  T. Hwa, P. Le Doussal, D. R. Nelson, and V. M. Vinokur,
Phys. Rev. Lett. {\bf 71}, 3545 (1993).

\bibitem{splay} L. Krusin-Elbaum, A. D. Marwick, R. Wheeler, C.
Feild, V. M. Vinokur, G. K. Leaf, and M. Palumbo, Phys. Rev. Lett.
{\bf 76}, 2563 (1996).

\bibitem{wengel}  C. Wengel and U. C. T\"{a}uber, Phys. Rev. B {\bf 58},
6565 (1998).

\bibitem{krauth}  W. Krauth, T. Trivedi, and D. Ceperley, Phys. Rev. Lett.
{\bf 67}, 2307 (1991).

\bibitem{sugano} R. Sugano, T. Onogi, K. Hirata, and M. Tachiki,
Phys. Rev. Lett. {\bf 80}, 2925 (1998).

\bibitem{sato}  Sato {\it et al.}, Phys. Rev. Lett. {\bf 79}, 3759 (1997);
Kosugi {\it et al.}, {\it ibid} {\bf 79}, 3763 (1997); Tsuchiya
{\it et al.}, Phys. Rev. B {\bf 59}, 11568 (1999).

\bibitem{morozov}  N. Morozov {\it et al.}, Phys. Rev. B {\bf 57}, R8146 (1998);
Phys. Rev. Lett. {\bf 82}, 1008 (1999).

\bibitem{vanderbeek}  N. Chikumoto {\it et al.}, Phys. Rev. B {\bf 57}, 14507
(1998); C. J. Van der Beek {\it et al.}, {\it ibid} {\bf 61}, 4259
(2000); K. Itaka {\it et al.}, preprint (2000).

\end{references}
\end{document}